\documentclass[10pt,twocolumn,twoside]{IEEEtran} 

\usepackage{cite}
\usepackage{amsmath,amssymb,amsfonts}
\usepackage{algorithmic}
\usepackage{graphicx}
\usepackage{textcomp}

\usepackage{mathdots}
\def\BibTeX{{\rm B\kern-.05em{\sc i\kern-.025em b}\kern-.08em
    T\kern-.1667em\lower.7ex\hbox{E}\kern-.125emX}}

\usepackage[table]{xcolor}    
\usepackage{array}  
\usepackage{cancel} 

\newtheorem{theorem}{Theorem}

\newtheorem{remark}{Remark}
\newtheorem{definition}{Definition}
\newtheorem{assumption}{Assumption}

\DeclareMathOperator*{\argmin}{argmin}

\begin{document}

\title{A Non-Conservative Stability Criterion for Networked Control Systems with time-varying Packet Delays}
\author{Martin Steinberger and Martin Horn
	
\thanks{Manuscript received ??, 2021;
M. Steinberger and M. Horn were supported by the LEAD project ``Dependable Internet of Things in Adverse Environments'' funded by Graz University of Technology. The financial support by the Christian Doppler Research Association, the Austrian Federal Ministry for Digital and Economic Affairs and the National Foundation for Research, Technology and Development is gratefully acknowledged. 
}
\thanks{M. Steinberger and M. Horn are with the Institute of Automation and Control, Graz University of Technology, 8010 Graz, Austria (e-mail: martin.steinberger@tugraz.at; martin.horn@tugraz.at).}
\thanks{M. Horn is with the Christian Doppler Laboratory for Model Based Control of Complex Test Bed Systems, Institute of Automation and Control, Graz University of Technology, 8010 Graz, Austria.}
}

\maketitle

\begin{abstract}
	A networked output feedback loop subject to packetized transmissions of the output signal is considered. Based on the small gain theorem, an easy-to-use stability criterion covering two important cases is presented. In the first case a packet numbering mechanism is employed whereas in the second case neither packet numbering nor synchronization between sender and receiver is assumed.
	The analysis makes use of acausal subsystems and deduces the optimal constant time delay that should be used in a nominal controller design such that additional packet delay variations introduced by the network are maximized. 
	A simulation example of a networked control system with a filtered Smith predictor  illustrates the application of the proposed criterion and compares the results to different approaches from literature.
\end{abstract}


\section{Introduction}

The most fundamental question in the design and analysis of networked control systems (NCS) is how to guarantee losed loop stability under the presence of network imperfections, as pointed out, e.g., in \cite{Park2018,Zhang2017,Gupta2010}. 
Stability criteria based on Linear Matrix Inequalities (LMIs) are widespread in literature. They follow different ideas to prove stability of networked loops for a variable time delay that should be as large as possible. In addition, the number of variables in the LMIs should be as small as possible in order to reduce the computational complexity, see, for example, \cite{Seuret2015,Li2011} and references therein.

Different control design methods rest on the stability analysis. In \cite{Cloosterman2010}, over-approximation techniques are used to design state controllers for loops with time-varying delays using LMI conditions.  
Sliding mode approaches are utilized in \cite{Ludwiger2019} together with a buffering mechanism to robustly stabilize spatially distributed networked feedback loops and render them insensitive to unknown bounded input disturbances. 
An alternative approach for the stability analysis is presented in \cite{Kao2004}. It extends the small gain theorem (SGT) for feedback loops with variable time delays.
Surprisingly, stability is not always explicitly considered as, e.g., in \cite{Batista2018}, where an adaptive Smith predictor is applied to control an optical oven over the internet that is subject to variable time delays.

It is crucial for the stability analysis of networked loops with time-varying delays to also consider the packetized nature of network transmissions. This means, that, e.g., the measurement data is sent in separate packets over a transmission channel. Due to the fact that each packet may experience a different packet delay, the used packet skipping and hold mechanisms at the receiver side have to be taken into account, see \cite{Steinberger2020_ifac}. This is an important step that is either implicitly considered as in \cite{Cloosterman2010}  or not included in the analysis as, e.g., in \cite{Li2011} and \cite{Kao2004}.

The authors of \cite{Steinberger2021_lcss} proposed a way how to include time-varying packet delays in the analysis by using a criterion based on the SGT. This work was extended in \cite{Steinberger2020_arxiv} to a network setup, where neither synchronization nor packet numbering is implemented. This allows to analyze feedback structures, where no packet reordering mechanism as, e.g., proposed in \cite{Liu2015} is utilized.
In \cite{Steinberger2020_arxiv}, a robust stability criterion is presented as well, which allows to account for uncertain plant models in feedback loops with variable packet delays.
However, the results from \cite{Steinberger2021_lcss} and \cite{Steinberger2020_arxiv} might be conservative. Consequently, the contributions of the present paper are:
\begin{enumerate}
	\item[(a)] The existing SGT-based approaches are enhanced to get a less conservative stability criterion. This yields a larger range for the admissible time-varying packet delays.
	
	\item[(b)] A splitting of the considered NCS into a causal and an acausal subsystem is used to minimize the effect of the uncertain variable delay. This analysis also yields an optimal constant time delay for a nominal controller design. 
	
	\item[(c)] The proposed computationally inexpensive criterion is applied to a networked loop consisting of an unstable plant and a filtered Smith predictor to underline the properties of the approach and compare the results to existing methods.	
\end{enumerate}

\emph{Notation:} Entire sequences are written as $(y_k) = (y_0, y_1, y_2, \ldots)$, one element is symbolized by $y_k$, where $k\in\mathbb{N}$ is the iteration index. $\big|\big|(y_k)\big|\big|_2$ is the 2-norm of sequence $(y_k)$. The z-transform of a sequence $(y_k)$ is denoted as $\tilde y(z)=\mathcal{Z}\left\{(y_k)\right\}$.
Discrete-time transfer functions $G(z)$ are written as functions of variable $z$. The infinity norm $\vert\vert G(z)\vert\vert_{\infty}$ is defined as the maximum of the corresponding magnitude plot of $G(z)$ that follows for $z = e^{j \omega h}$ and frequencies $\omega\in \left[0,\pi/h\right)$, where $h$ symbolizes the constant sampling time.

\section{Problem Statement}

\begin{figure}[!t]\centering
	\includegraphics[width=0.8\columnwidth]{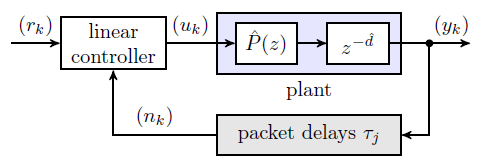}
	\caption{Feedback loop that is closed by means of a communication network with variable time delays $\tau_j$ for the transmitted packets $j$.} 
	\label{fig:unity_feedback_2delays}
\end{figure}

The considered feedback loop consists of a plant, a packetized transmission network and a linear controller, see Fig.~\ref{fig:unity_feedback_2delays}. The plant with input sequence $(u_k)$ and output sequence $(y_k)$ is given as
\begin{equation}\label{eq:Pz}
	P(z) = \hat P(z) z^{-\hat d} \, ,
\end{equation}
where $\hat P(z)$ is a proper nominal discrete-time transfer function and $0 \le \hat d  \in\mathbb{N}$ represents a nominal plant delay .
As a controller, one can, e.g., use $u(z) = C(z) \left(r(z) - n(z) \right)$ as in \cite{Steinberger2021_lcss} to get a unity feedback loop, or filtered Smith predictors as presented in \cite{Steinberger2020_arxiv}.

\begin{assumption}[Network delays]\label{as:delay}
	The elements of output sequence $(y_k)$ are transmitted in separate packets $j$ that are subject to individual bounded packet delays $\tau_j$ so that
	\begin{equation}\label{eq:NW_delay}
		0 \le \underline\tau_N \le \tau_j \le \bar\tau_N 
	\end{equation}
	with  $0 \le \underline\tau_N <\bar\tau_N$ and $\underline\tau_N, \tau_j, \bar\tau_N\in \mathbb{N}$.	
\end{assumption}
Please note that there are no further assumptions on the delay distribution nor on the maximal admissible change rate of two subsequent packet delays. The elements of output $y_k$ are sent in individual packets $j$ via the transmission channel to a receiver at the controller side (see Fig.~\ref{fig:unity_feedback_2delays}) and may arrive at the same time instant or out of order due to the time-varying transmission delays. Hence, it is important to specify the used protocols, i.e. the packet selection and skipping mechanism as well as the hold mechanism that is active whenever no packet arrives at the receiver side. Two out of the three different protocols, formally introduced in \cite{Steinberger2021_lcss} and \cite{Steinberger2020_arxiv}, are considered in this work\footnote{For the sake of compatibility, the same notation as in \cite{Steinberger2020_arxiv} is used in the present paper.}:

Protocol $\mathcal{P}_1$: The most recent packet is used, if more packets are available at the same time. Packets are skipped, if they arrive after a more recent packet has already been received. This requires either a synchronization of sender and receiver or at least a numbering of subsequently sent packets.

Protocol $\mathcal{P}_3$: Neither synchronization nor packet numbering is used. This represents the most reduced network transmission approach with the least possible overhead. Consequently, it constitutes the worst case for stability analysis as the order of arriving packets is unknown and any arriving packet may be selected at the receiver side.

Both protocols make use of a zero order hold mechanism on the receiver side. Compared to $\mathcal{P}_1$ and $\mathcal{P}_3$, protocol $\mathcal{P}_2$ presented in \cite{Steinberger2021_lcss} is an intermediate case and is not considered here. In principle, the proposed stability criterion can also be formulated for  $\mathcal{P}_2$.

Combining plant delay \eqref{eq:Pz} and network delay \eqref{eq:NW_delay} results in a  constant time delay $\hat d +\underline\tau_N$ and a time-varying delay that is bounded such that 
\begin{equation}\label{eq:tau_hat_N}
	0 \le \tau_j - \underline\tau_N \le \bar\tau_N -\underline\tau_N = \hat\tau_N \, .
\end{equation} 
For the nominal controller design and stability analysis presented below, an additional constant acausal time delay $0 \le  \tau_A \in \mathbb{N}$ is introduced, yielding a modified constant time delay
\begin{equation}\label{eq:tau_nominal_ac}
	\hat\tau = \hat d +\underline\tau_N + \tau_A
\end{equation}
and a (usually) acausal time-varying delay
\begin{equation}\label{eq:tau_bounds_ac}
	-\tau_A \le \tau_j -  \underline\tau_N - \tau_A \le \hat\tau_N-\tau_A 
\end{equation}
for the overall delay of plant and communication network. The specific choice of $\tau_A=0$ constitutes the causal case as in \cite{Steinberger2021_lcss,Steinberger2020_arxiv}.

The goals of the present papers are: (i) derive less conservative stability conditions for protocols $\mathcal{P}_1$ and $\mathcal{P}_3$ compared to \cite{Steinberger2021_lcss} and \cite{Steinberger2020_arxiv}; (ii) find an optimal choice $\tau_A^\ast$ for the acausal delay resulting in an increase of the admissible time-varying network delay $\hat\tau_N$ maintaining finite gain $\ell_2$ stability of the closed loop.

\section{Stability Criterion}
\label{sec:stabiliy}

\begin{figure}[!t]\centering
	\includegraphics[width=\columnwidth]{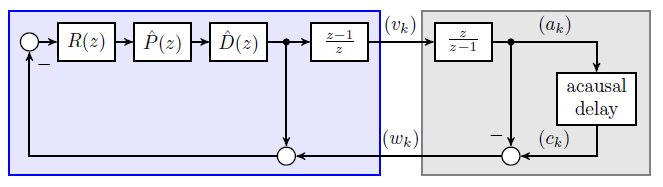}
	\caption{Restructured feedback loop that is used for the stability analysis: nominal part (blue); uncertainty due to the (acausal) time-varying delays (gray).}
	\label{fig:SGT}
\end{figure}

This section introduced the proposed stability criterion that rests on the separation of the original structure, presented in Fig~\ref{fig:unity_feedback_2delays}, into a nominal part (blue in Fig.~\ref{fig:SGT}) and a part characterizing the uncertainty due to the time-varying packet delays (gray block in Fig.~\ref{fig:SGT}).
In addition, the reference input $(r_k)$ is zero and the remaining linear controller in Fig.~\ref{fig:unity_feedback_2delays} is represented by transfer function $R(z)$
with input sequence $(n_k)$, delayed measurement sequence $(y_k)$ and output sequence $(u_k)$. A discrete-time integrator and differentiator are utilized to avoid issues for the case, where the nominal controller $R(z)$ is designed to achieve a dc-gain equal to one for the nominal loop, see  \cite{Steinberger2021_lcss} for details. The effect of the known constant time delay $\hat\tau$ \eqref{eq:tau_nominal_ac} is included in the nominal (blue) part via transfer function
\begin{equation}\label{eq:Dz}
	\hat D(z) = z^{-\hat\tau} \ .
\end{equation}
Note that the acausal delay $\tau_A$ also appears in \eqref{eq:Dz} because of \eqref{eq:tau_nominal_ac}. Before the following theorems are stated, some definitions are recalled. For more details see, e.g., \cite{Vidyasagar2002}.
\begin{definition}[Truncation of a sequence]\label{def:truncation}
	The truncation of sequence $(f_k):[a,\infty)\rightarrow\mathbb{R}$ at a finite time $T$ is defined as
	\begin{equation}
		\big(f_k\big)_T = \begin{cases}
		    0   & \forall t \ \text{if}\ T< a\\
			f_k & a \le k \le T \\
			0   & k > T \ge a 
		\end{cases} \, .
	\end{equation}
\end{definition}
\begin{definition}[Extended $\ell_2$ space]
	A sequence $(f_k)$ belongs to $\ell_2 [a,\infty)$ if condition $\sum_{k=a}^{\infty} \big|f_k\big|^2 < \infty$ holds. The sequence $(f_k)\in\ell_{2e} [a,\infty)$, if its truncation belongs to $\ell_2 [a,\infty)$, i.e. $\big(f_k\big)_T\in\ell_{2} [a,\infty)$ for all $T$.
\end{definition}
\begin{definition}[Finite gain $\ell_2$ stability]
	\label{def:finite_lp_stability}
	A mapping $(y_k)=\mathcal{M}\{(u_k)\}: \ell_{2e} [a,\infty) \mapsto\ell_{2e} [a,\infty)$ is finite gain $\ell_2$ stable if there exist constants $\alpha>0$ and $\beta>0$ such that
	\begin{equation}
		\big|\big|\big(y_k\big)\big|\big|_2 \le \alpha 	\big|\big|\big(u_k\big)\big|\big|_2 + \beta 
	\end{equation}
	for $(u_k)\in\ell_2 [a,\infty)$ and $(y_k)\in\ell_2 [a,\infty)$, i.e. there is an affine bound of the norm of output sequence $(y_k)$. The constant $\alpha$ is referred to as the finite $\ell_2$ gain.
\end{definition}
\begin{definition}[Causality]\label{def:causality}
	A mapping $(y_k)=\mathcal{M}\{(u_k)\}: \ell_{2e} [a,\infty) \mapsto\ell_{2e} [a,\infty)$ is causal if
	\begin{equation}
		(y_k)_T = \big(\mathcal{M}\big\{(u_k)\big\}\big)_T = \big(\mathcal{M}\big\{(u_k)_T\big\}\big)_T	
	\end{equation}
	for all finite $T>0$ and $(u_k)\in\ell_{2e} [a,\infty)$.
\end{definition}
Please note that there are definitions of the finite $\ell_2$ gain that are different to \cite{Vidyasagar2002}. In \cite{Sastry1999}, the $\ell_2$ gain is defined for causal mappings, i.e. it is a combination of Definition~\ref{def:finite_lp_stability} and \ref{def:causality} using the truncation operator such that $||(y_k)_T||_2 \le \alpha 	||(u_k)_T||_2 + \beta $ for $(u_k)\in\ell_{2e} [a,\infty)$.

\begin{figure}[!t]\centering
	\medskip
	\includegraphics[width=0.8\columnwidth]{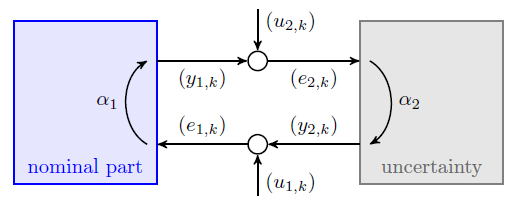}
	\caption{Feedback loop for the formulation of the discrete-time small gain theorem. Either the nominal subsystem or the uncertainty might be acausal.}
	\label{fig:SGT_acausal}
\end{figure}
Based on \cite{Uenal2008} where an SGT for continuous-time systems with acausal subsystems is introduced, we can now state the discrete-time SGT for the feedback loop shown in Fig.~\ref{fig:SGT_acausal} with $(u_{1,k}), (u_{2,k})\in\ell_{2} [a,\infty)$ and $(e_{1,k}), (e_{2,k}), (y_{1,k}), (y_{2,k})\in\ell_{2e} [a,\infty)$. The nominal part and the uncertainty are characterized by mappings $(y_{1,k})=\mathcal{M}_1\{(e_{1,k})\}$ and $(y_{2,k})=\mathcal{M}_2\{(e_{2,k})\}$ respectively. They are assumed to be finite gain $\ell_2$ stable, i.e.
\begin{subequations}\label{eq:gains}
	\begin{align}
	\big|\big| (y_{1,k}) \big|\big| &\le \alpha_1 \big|\big| (e_{1,k}) \big|\big| + \beta_1 \quad  \forall (e_{1,k})\in\ell_{2} [a,\infty) \ , \label{eq:gain1}\\
	\big|\big| (y_{2,k}) \big|\big| &\le \alpha_2 \big|\big| (e_{2,k}) \big|\big| + \beta_2 \quad  \forall (e_{2,k})\in\ell_{2} [a,\infty) \ .\label{eq:gain2}
	\end{align}
\end{subequations}
If both mappings are causal, the classical SGT as stated in \cite{Vidyasagar2002}, \cite{Sastry1999} can be applied. However, for the stability analysis of networked control systems as considered in this paper, it will turn out to be beneficial that one subsystem is acausal. 

\begin{theorem}[Acausal SGT for discrete-time systems]\label{th:SGT_general}
	Consider the feedback loop shown Fig.~\ref{fig:SGT_acausal} with finite $\ell_2$ gains \eqref{eq:gains}. Let 
	\begin{equation} \label{eq:SGT_general_alphas}
		\alpha_1 \alpha_2 <1
	\end{equation}
	and conditions 
	\begin{subequations}
	\begin{align} 
		\Big( \mathcal{M}_1\Big\{ \mathcal{M}_2\big\{(e_{2,k})\big\} \Big\}\Big)_T &= \Big( \mathcal{M}_1\Big\{ \mathcal{M}_2\big\{(e_{2,k})_T\big\} \Big\}\Big)_T \label{eq:SGT_general_causal_cond1}\\
		\Big( \mathcal{M}_2\Big\{ \mathcal{M}_1\big\{(e_{1,k})\big\} \Big\}\Big)_T &= \Big( \mathcal{M}_2\Big\{ \mathcal{M}_1\big\{(e_{1,k})_T\big\} \Big\}\Big)_T \label{eq:SGT_general_causal_cond2}\\
		\Big\vert\Big\vert \Big( \mathcal{M}_\nu \big\{ \big(e_{\nu,k}^{(1)}\big) &+ \big(e_{\nu,k}^{(2)}\big) \big\} \Big)_T \Big\vert\Big\vert_2 \le \label{eq:SGT_general_linear_cond}\\
		\Big\vert\Big\vert \Big( \mathcal{M}_\nu \big\{ \big(e_{\nu,k}^{(1)}\big)_T \big\} \Big)_T \Big\vert\Big\vert_2 &+ 
		\Big\vert\Big\vert \Big( \mathcal{M}_\nu \big\{ \big(e_{\nu,k}^{(2)}\big)_T \big\} \Big)_T \Big\vert\Big\vert_2 , \nonumber 
	\end{align}
	\end{subequations}
	hold for all $(e_{\nu,k}), (e_{\nu,k}^{(1)}), (e_{\nu,k}^{(2)}) \in\ell_{2e}[a,\infty)$ and $\nu\in\{1,2\}$.\\
	Then, the closed loop system is finite gain $\ell_2$ stable.
\end{theorem}
\begin{IEEEproof}
	The proof directly follows as discrete-time counterpart of the continuous-time version from \cite{Uenal2008} evaluated for $n_1=n_2=1$ and  $a_1=a_2=a$.
\end{IEEEproof}

\begin{remark}\label{re:linearity}
	Conditions \eqref{eq:SGT_general_causal_cond1} and \eqref{eq:SGT_general_causal_cond2} ensure, that the cascade connection of $\mathcal{M}_1$ and $\mathcal{M}_2$ (and vice versa) are causal. For example, $(y_{1,k}) = \mathcal{M}_1\{(e_{1,k})\} = \mathcal{M}_1\{(u_{1,k})+(y_{2,k})\} = \mathcal{M}_1\{(u_{1,k})+\mathcal{M}_2\{(e_{2,k})\}$ has to be causal according to Definition~\ref{def:causality} for $(u_{1,k})$ identically to zero as claimed in \eqref{eq:SGT_general_causal_cond1}. 
\end{remark}
\begin{remark}
	Please note that condition \eqref{eq:SGT_general_linear_cond} is automatically fulfilled if both subsystems in Fig.~\ref{fig:SGT_acausal} are linear.
\end{remark}

Theorem \ref{th:SGT_general} establishes the basis for the proposed stability criterion for feedback loops as depicted in Fig.~\ref{fig:unity_feedback_2delays}.
The nominal part with input $(w_k)$ and output $(v_k)$ in Fig.~\ref{fig:SGT} can be described by transfer function
\begin{equation}\label{eq:Mz}
	M(z) = \frac{-R(z) \hat	P(z) \hat D(z)}{1+R(z) \hat P(z) \hat D(z)} \frac{(z-1)}{z} 
\end{equation}
with the corresponding finite $\ell_2$ gain of $\alpha_1 = || M(z) ||_{\infty}$. In contrast, it is more challenging to find the $\ell_2$ gain $\alpha_2=\alpha$ of the uncertainty in Fig.~\ref{fig:SGT}. This is because (acausal) time-varying packet delays in combination with protocol $\mathcal{P}_1$ or $\mathcal{P}_3$ have to be considered. Hence, we follow the general procedure proposed in \cite{Steinberger2021_lcss}, \cite{Steinberger2020_arxiv} and extend it to include the acausal delay $\tau_A$. To find the $\ell_2$ gain $\alpha$ associated with the uncertainty, on has to maximize the norm of output $w_k = c_k - a_k$, where $a_k$ is the summation of input sequence $(v_k)$. Since the uncertainty is linear but time-varying, one can use input sequence $(v_k)_T = \big(v_0, v_1, v_2, \ldots, v_T, 0, \ldots\big) = \big(\bar v, \bar v, \bar v, \ldots, \bar v, 0,  \ldots\big)$ with constant $\bar v$ and truncation time $T>0$. As a result, sequence $(a_k) = \big(a_0, a_1, a_2, \ldots, a_T, a_{T+1}, \ldots\big) 	= \big(\bar v, 2\bar v, 3\bar v, \ldots, (T+1)\bar v, (T+1)\bar v, \ldots\big)$ is transmitted and experiences time-varying delays bounded as stated in \eqref{eq:tau_bounds_ac}. 
\begin{figure}[!t]\centering
	\includegraphics[width=\columnwidth]{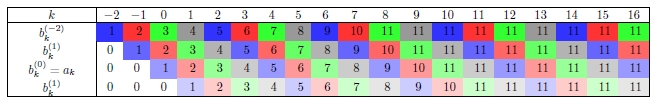}
	\caption{Packets on the way for $\hat\tau_N=3$, $\bar v=1$ and $\tau_A=2$. Darker colors represent more recent packets.}
	\label{fig:packets_on_the_fly}
\end{figure}
Figure~\ref{fig:packets_on_the_fly} shows all possible packets on the way for $\hat\tau_N=3$ and an acausal delay of $\tau_A=2$. This means that, e.g., the packet containing $a_5=6$ might arrive at the receiver side at time instants $k\in\{3,4,5,6\}$ as can be seen in Fig.~\ref{fig:packets_on_the_fly} (red packets labeled with number $6$). The set of all packets that might arrive at time instant $k$ is $\big\{b_k^{(-\tau_A)}, b_k^{(-\tau_A+1)}, \ldots, b_k^{(\hat\tau_N-\tau_A)}\big\}$.
Depending on the actual protocol, e.g., the most recent packet is selected as $c_k$ and further used. Consequently, the $\ell_2$ gain of the uncertainty is
\begin{equation}\label{eq:alpha}
	\alpha = \sup_{T>0} \alpha_T \quad\text{with}\quad \alpha_T = \sqrt{\frac{\big|\big|(w_k)_T\big|\big|_2^2}{\big|\big|(v_k)_T\big|\big|_2^2}},
\end{equation}
and $||(v_k)_T||_2^2=(1+T) \bar v^2$. Please note that $\alpha$ depends on $\tau_A$, which will be chosen to minimize the finite $\ell_2$ gain.

Table~\ref{tab:P3_optimal_values} shows the optimal acausal delays $\tau_A^\ast$ that yield a minimization of the related gain  for different maximal delays $\hat\tau_N$ and protocol $\mathcal{P}_3$. Details on the calculations can be found in Appendix \ref{sec:proof}. With this, we are able to state the main stability criterion for the considered NCS. 

\begin{table}\centering
	\caption{Optimal acausal delays $\tau_A^\ast$ and related $\ell_2$ gains $\alpha^\ast = g(\tau_A^\ast, \hat\tau_N)$ for different maximal admissible variable network delays $\hat\tau_N$ and protocol $\mathcal{P}_3$.}
	\label{tab:P3_optimal_values}
	\includegraphics[width=\columnwidth]{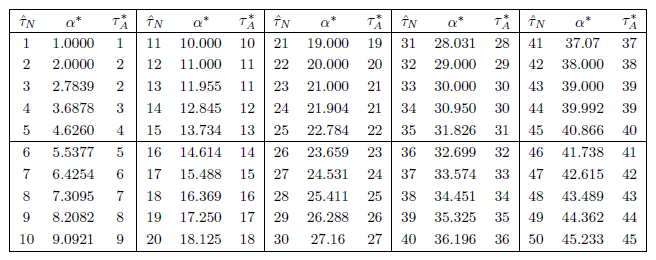}
\end{table}
\begin{theorem}[Stability criterion for NCS using acausal subsystems]\label{th:NCS_stability}
	Consider the networked feedback loop shown in Fig.\ref{fig:unity_feedback_2delays} with plant \eqref{eq:Pz} and a packetized transmission network subject to network delays \eqref{eq:NW_delay}. Let the linear controller $u(z)=-R(z) n(z)$ for $(r_k)=0\ \forall k$ be designed for a nominal delay $\hat\tau = \hat d + \underline\tau_N +\tau_A^\ast$ and suppose that acausal delay $\tau_A^\ast$ and the corresponding finite $\ell_2$ gain $\alpha^\ast$ are given such that
	\begin{enumerate}
		\item[(a)] for protocol $\mathcal{P}_1$ (skip old packets, take newest if more packets are available):
		\begin{subequations}\label{eq:th_tauA_alpha_P1}
		\begin{align}
			\tau_A^\ast &=\bigg\lceil \frac{\bar\tau_N+\underline\tau_N}{2} \bigg\rceil \quad\text{or}\quad \tau_A^\ast = \bigg\lfloor \frac{\bar\tau_N+\underline\tau_N}{2} \bigg\rfloor \, ,\label{eq:th_tauA_P1}\\
			\alpha^\ast &= \alpha_{\mathcal{P}_1}^\ast = \max\big\{\tau_A^\ast, \hat\tau_N-\tau_A^\ast \big\} \label{eq:th_alpha_P1}
		\end{align}
		\end{subequations}
		
		\item[(b)] for protocol $\mathcal{P}_3$ (no numbering nor synchronization, worst case)
		\begin{equation}
			\alpha^\ast = \alpha_{\mathcal{P}_3}^\ast = g\left(\tau_A^\ast, \hat\tau_N\right)
			\label{eq:th_alpha_P3}
		\end{equation}
		in accordance with the Table~\ref{tab:P3_optimal_values}.
	\end{enumerate}

	The feedback loop is finite gain $\ell_2$ stable for all bounded time-varying packet delays $0 \le \underline\tau_N \le \tau_j \le \bar\tau_N$, if	
	\begin{equation}
		\Big|\Big| M(z)\Big|\Big|_{\infty} \alpha^\ast = \Bigg|\Bigg| \frac{R(z) \hat 	P(z)}{1+R(z) \hat P(z) \hat D(z)} \frac{(z-1)}{z}\Bigg|\Bigg|_{\infty} \alpha^\ast <1
		\label{eq:th_cond}
	\end{equation}
	is fulfilled. Constant $\tau_A^\ast$ is the optimal acausal time delay that minimizes the finite $\ell_2$ gain $\alpha^\ast$ for a given $\hat\tau_N$.
\end{theorem}

\begin{IEEEproof}
	A proof is given in Appendix \ref{sec:proof}.
\end{IEEEproof}

\begin{remark}\label{re:overest}
	Note that relation \eqref{eq:th_alpha_P3} cannot be stated explicitly, as it is also visible in  Table~\ref{tab:P3_optimal_values}. The steps, how to numerically find optimal values $\tau_A^\ast$ and $\alpha^\ast$, are detailed in Appendix~\ref{sec:proof}. However, it is possible to over-estimate the non-linear characteristics in Table~\ref{tab:P3_optimal_values} using results from protocol $\mathcal{P}_1$ so that	
	\begin{equation}\label{eq:alpha_overest}
		\tau_A = \hat\tau_N \qquad\text{and}\qquad \alpha = \alpha_{\mathcal{P}_1}\big\vert_{\tau_A=\hat\tau_N} = \hat\tau_N
	\end{equation}
	hold for protocol $\mathcal{P}_3$ as depicted in Fig.~\ref{fig:P3_tauA_optimal}.
	\begin{figure}[!t]\centering
		\includegraphics[width=\columnwidth]{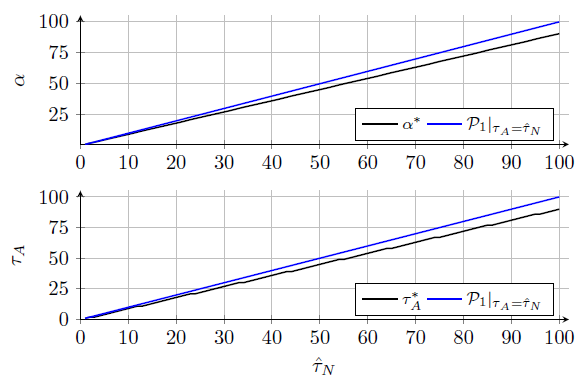}
		\caption{Optimal acausal time delay $\tau_A^\ast$ and gain $\alpha^\ast$ for protocol $\mathcal{P}_3$ (black) and a possible over-estimation (blue).}
		\label{fig:P3_tauA_optimal}
	\end{figure}
	The specific patterns of packet delays leading to the $\ell_2$ gains for both protocols are presented in the Appendix.
\end{remark}
\begin{remark}
	For the causal case, i.e. for $\tau_A=0$, one obtains the same conditions as (i) in \cite{Steinberger2021_lcss} for a unity feedback loop closed via a network channel, where $R(z)=C(z)$,  $\hat d = 0$, $\underline \tau_N=0$ and, as a result, $\hat D(z)=1$ (no delay in plant); (ii) in \cite{Steinberger2020_arxiv} for a NCS with a filtered Smith predictor, $R(z)=\frac{C(z) F(z)}{1+C(z)H(z)}$, $H(z)=\hat P(z) \big(1-\hat D(z) F(z)\big)$ and $\hat d \ge 0$, $\underline \tau_N\ge0$, which implies $\hat D(z)=z^{-\hat\tau}=z^{-\hat d -\underline \tau_N}$.
\end{remark}

\begin{remark}
	Theorem~\ref{th:NCS_stability} can be extended in a straightforward way to uncertain plant models $P(z) = \hat P(z) z^{-\hat d} \big(1+\delta P\big)$ as presented in \cite{Steinberger2020_arxiv} for the causal case.
\end{remark}

\section{Simulation Example}

The application of Theorem~\ref{th:NCS_stability} is shown for an NCS consisting of a filtered Smith predictor and the unstable plant
\begin{equation} \label{eq:ex_plant}
   P(z) = \frac{0.0051271}{z-1.051} z^{-5}
\end{equation}
as used in \cite{Steinberger2020_arxiv} for the causal case with sampling time $h=1$. The nominal controller $C(z)$
with prefilter $V(z)$ in the form of $u(z) = C(z) \left(V(z) r(z) - n(z) \right)$ is designed such that the delay-free, nominal closed loop has poles at $p_1 = p_2 = 0.95$ and the overshoot during set-point changes is reduced by $V(z)$. This results in, transfer functions
\begin{equation}
	C(z) = \frac{29.504 (z-0.9835)}{z-1} \, ,\ V(z) = \frac{0.041317 (z-0.6)}{(z-0.9835)}.
\end{equation}
The pole of filter transfer function $F(z)$ for the predictor is fixed at $p=0.95$, see  \cite{Steinberger2020_arxiv} for details.
Since a filtered Smith predictor structure is considered, the nominal design does not depend on $\tau_A$. However, $\tau_A$ is taken into account in the overall constant delay \eqref{eq:tau_nominal_ac} for the predictor.

Evaluation of the LMI conditions stated in \cite{Steinberger2020_arxiv} that are based on \cite{Li2011}, yields either $5$ or $1$ for the maximal admissible time-varying delay $\hat\tau_N$, see also Table~\ref{tab:example_comp}. However, as shown in \cite{Steinberger2020_arxiv}, there are sequences of packet delays that already lead to instability for $\hat\tau_N=4$. This is because the packetized character of the transmissions is not explicitly considered in the analysis. This issue is overcome by using Theorem~2 from \cite{Steinberger2020_arxiv}, where the SGT is utilized for the causal case. The maximum achievable $\hat\tau_N$ is $4$ for protocol $\mathcal{P}_1$ and $2$ for $\mathcal{P}_3$.
\begin{table}\centering
	\caption{Example: maximal admissible variable time delay $\hat\tau_N$ for conventional LMI approaches and SGT approaches that include the packetized nature of network transmissions.}
	\label{tab:example_comp}
	\includegraphics[width=0.8\columnwidth]{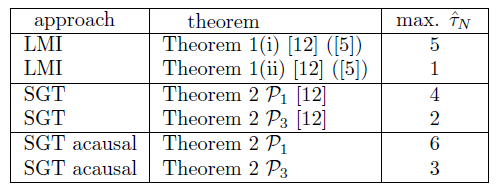}
\end{table}

\begin{figure}[!t]\centering
	\includegraphics[width=\columnwidth]{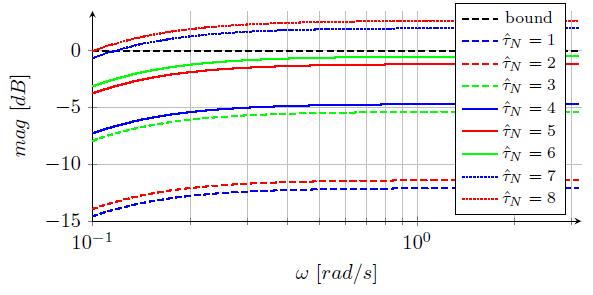}
	\caption{Example: magnitude plots for $M(z) \alpha^\ast$ using protocol $\mathcal{P}_1$ and different maximal network delays $\hat\tau_N$.}
	\label{fig:bode_P1}
\end{figure}

\begin{figure}[!t]\centering
	\includegraphics[width=\columnwidth]{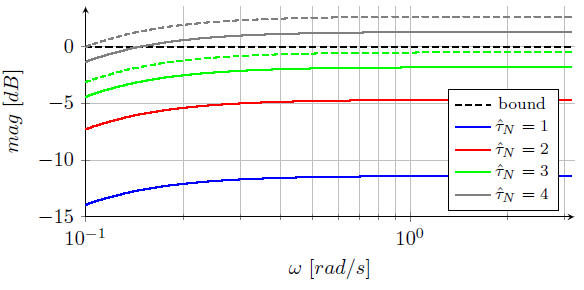}
	\caption{Example: magnitude plots for $M(z) \alpha^\ast$ using protocol $\mathcal{P}_3$ and different maximal network delays $\hat\tau_N$. Dotted lines represent results obtained by using the over-estimated values \eqref{eq:alpha_overest}.}
	\label{fig:bode_P3}
\end{figure}

Next, Theorem~\ref{th:NCS_stability} is evaluated for the example. Conditions \eqref{eq:th_tauA_alpha_P1}, and \eqref{eq:th_alpha_P3} are evaluated for $\underline\tau_N=0$ and different $\hat\tau_N$. The optimal acausal delay $\tau_A^\ast$ is used in the predictor and relation \eqref{eq:th_cond} is checked by using bode magnitude plots for $M(z) \alpha^\ast$ as shown in Fig.~\ref{fig:bode_P1} and \ref{fig:bode_P3}.
Figure~\ref{fig:bode_P1} reveals that one can show stability for a maximal value of $\hat\tau_N=6$ with the proposed theorem for $\mathcal{P}_1$, which is larger than using the LMI-based approach.

A maximum $\hat\tau_N$ of $3$ is achieved for protocol $\mathcal{P}_3$ as depicted in Fig.~\ref{fig:bode_P3}. Please note that one can find delay patterns \cite{Steinberger2020_arxiv} for $\hat\tau_N=4$ that lead to instability. Hence, the upper bound provided by Theorem~\ref{th:NCS_stability} equals the largest achievable value for the admissible time-varying delay. Over-estimation \eqref{eq:alpha_overest} for $\mathcal{P}_3$ will often lead to a smaller admissible $\hat\tau_N$, which is not the case for the presented example, see green dashed line in Fig.~\ref{fig:bode_P3}. Table~\ref{tab:example_comp} presents a comparison of the results for all considered approaches for the simulation example.

\section{Conclusion}

A stability criterion based on a SGT for feedback loops with acausal subsystems is proposed that allows to achieve less conservative results compared to existing approaches in literature. It allows to show stability for a larger admissible variable time delay introduced by a communication network. Since packets containing measurement data are sent in separate packets over the transmission channel, it is vital to also include the packetized nature of the transmissions. This is possible by using an extension of the SGT for discrete-time systems (Theorem~\ref{th:SGT_general}), where one subsystem might be acausal. The acausality of one subsystem is exploited to maximize the achievable delay $\hat\tau_N$ for different protocols (Theorem~\ref{th:NCS_stability}). Additionally, the proposed approach yields an optimal acausal delay $\tau_A^\ast$ that is used to design the nominal controller for the constant delay $\hat d +\underline\tau_N + \tau_A^\ast$.

A simulation example for a NSC with filtered Smith predictor provides insights to the application of this easy-to-use criterion for networked loops and compares the results to LMI-based solutions, where the packet selection and hold mechanisms at the receiver side cannot be directly included.

\appendices
\section{Proof of Theorem~\ref{th:NCS_stability}}
\label{sec:proof}

The proof is based on the application of Theorem~\ref{th:SGT_general} to the feedback structure shown in Fig.~\ref{fig:SGT}. Hence, the gain of the nominal part and the uncertainty due to the time-varying delays are considered. Gain $\alpha_1$, as defined in \eqref{eq:gain1}, is given by 
$\alpha_1 = || M(z)||_{\infty}$, 
where \eqref{eq:Mz} is used for the description of the nominal part. Condition \eqref{eq:SGT_general_linear_cond} is fulfilled for the considered NCS because both subsystems (nominal part and uncertainty) are linear, see also Remark~\ref{re:linearity}. Since Fig.~\ref{fig:unity_feedback_2delays} is equivalent to Fig.~\ref{fig:SGT} for $(r_k)=(0, 0, \ldots)$, the cascade connections of both subsystems are causal, as claimed by \eqref{eq:SGT_general_causal_cond1} and \eqref{eq:SGT_general_causal_cond2}. Consequently, only the calculation of the finite $\ell_2$ gain $\alpha^\ast=\alpha_2$ for the uncertainty remains to show \eqref{eq:th_alpha_P1}, \eqref{eq:th_alpha_P3}, \eqref{eq:th_cond} in Theorem~\ref{th:NCS_stability}. This is done below for the gray subsystem in Fig.~\ref{fig:SGT} with input $v_k$ (with $\bar v=1$) and output $w_k$ considering protocols $\mathcal{P}_1$ and $\mathcal{P}_3$.

\subsection{Protocol $\mathcal{P}_1$}

\begin{figure}[!t]\centering
	\includegraphics[width=\columnwidth]{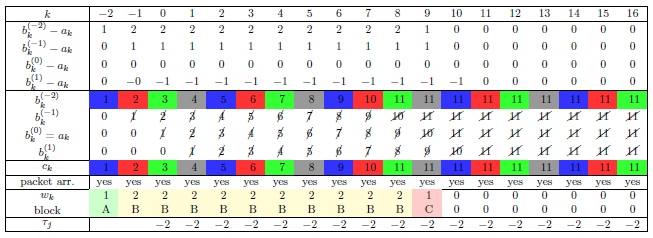}
	\caption{Packet pattern and inner signals of the uncertainty for $\mathcal{P}_1$, $\hat\tau_N=3$, $T=10$ and $\tau_A=2$}
	\label{fig:P1_tauMax3_-2_1}
\end{figure}

\begin{figure}[!t]\centering
		\includegraphics[width=\columnwidth]{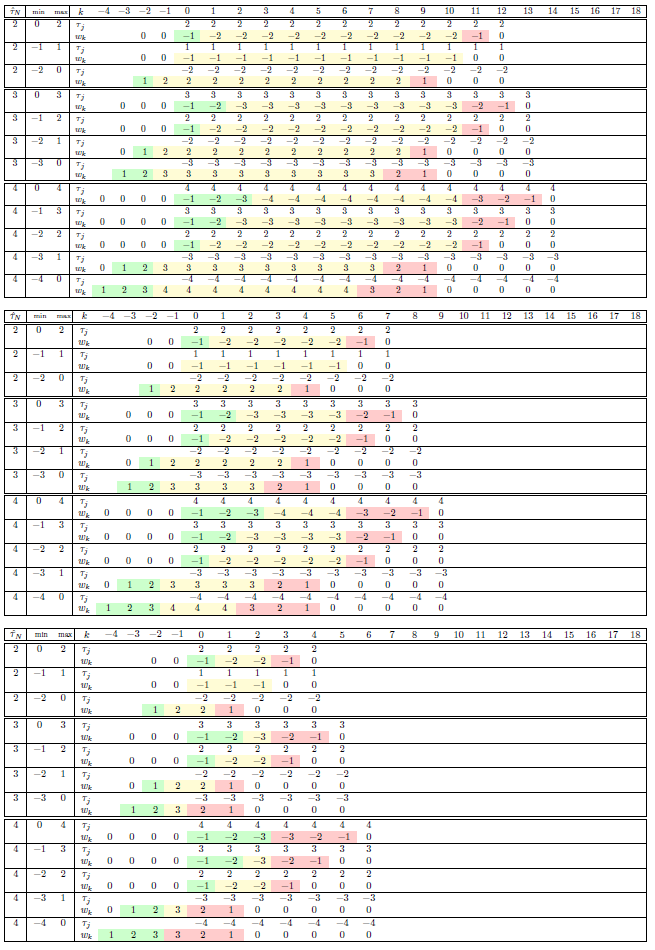}
	\caption{Resulting delay patterns $(\tau_j)$ and worst case output sequences $(w_k)$ for $\mathcal{P}_1$, $\hat\tau_N\in\{2,3,4\}$, $T=10$ (top), $T=5$ (middle), and $T=2$ (bottom).}
	\label{fig:P1_T10_T2} 
\end{figure}

Figure~\ref{fig:P1_tauMax3_-2_1} shows all inner signals as well as the output $w_k$ of the uncertainty for a maximal admissible variable time delay $\hat\tau_N=3$ and an additional acausal delay $\tau_A=2$. The input sequence $(v_k)$ and $(a_k)$, are truncated at $T=10$, as shown in Section~\ref{sec:stabiliy}.
To maximize the norm of $w_k$, all possible packets on the way $\big\{b_k^{(-\tau_A)}, b_k^{(-\tau_A+1)}, \ldots, b_k^{(\hat\tau_N-\tau_A)}\big\}$ (see also Fig.~\ref{fig:packets_on_the_fly}) are considered. For example at time instant $k=0$, the packet containing $3$ yields the largest absolute value for $w_k = b_k^{(-2)}-a_k$. Consequently, $a_2=3$ has to arrive at time instant $k=0$, i.e. two time instants before it is sent. Its corresponding packet delay is $\tau_3=-2$.
In addition, one has to take into account the fact, that each packet can only be received once at the receiver side and, as defined for protocol $\mathcal{P}_1$, the most recent packet is chosen if more packets are available. The packet received at the last time instant, i.e. $c_{k-1}$, is used if no packet is received at instant $k$.
Figure~\ref{fig:P1_tauMax3_-2_1} presents the resulting packet pattern that results in a maximization of the norm of $(w_k)$. The worst case delay pattern is $\tau_j=-2$ for all $j$.
Figure~\ref{fig:P1_T10_T2} extends the analysis to different $\hat\tau_N$, $T$ and acausal delays $\tau_A$, where also the minimal delay $-\tau_A$ and maximal delay $\hat\tau_N-\tau_A$ are indicated.
All resulting delay patterns and output sequences show a similar structure as for the causal case \cite{Steinberger2021_lcss}, with the important differences that: (i) a maximal relative delay
\begin{equation}\label{eq:tau_bar}
	\bar\tau = \max\big\{-\tau_A, \hat\tau_N-\tau_A \big\}
\end{equation}
is relevant; (ii) the norm of the output sequence for the introductory part A (green), main part B (yellow) and remaining part C (red) is given by $||(w_k)_T||_2^2 = A + B + C = 2A + B$,
\begin{align}\label{eq:wk_P1}
	\big|\big|(w_k)_T\big|\big|_2^2 = 2 \sum_{i=1}^{\bar\tau -1} i^2 \bar v^2 + \left(T-\bar\tau+2\right) \bar\tau^2 \bar v^2
\end{align}
for $T-\bar\tau+2 \ge 0$, i.e. $T \ge \bar\tau-2$. 

\begin{figure}[!t]\centering
	\includegraphics[width=\columnwidth]{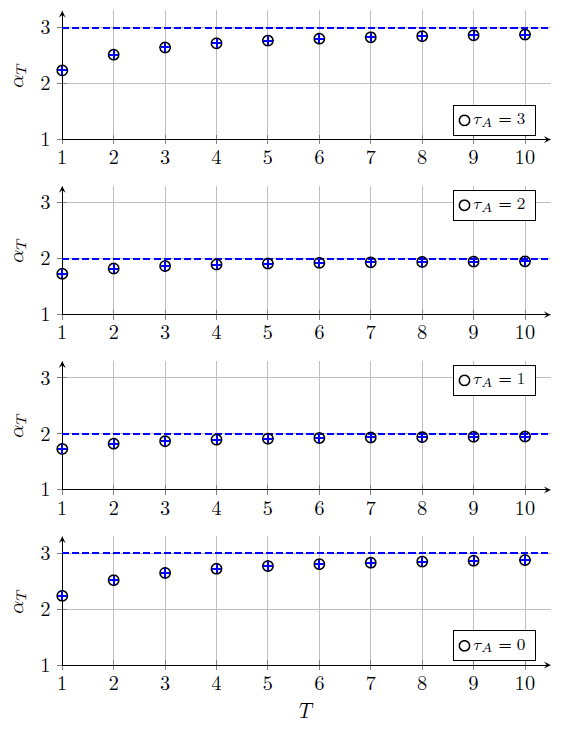}
	\caption{Gains as a function of truncation point $T$ and the acausal delay $\tau_A$ for $\mathcal{P}_1$ and $\hat\tau_N=3$.}
	\label{fig:alphaT_T_P1_tauMax_3}
\end{figure}

\begin{figure}[!t]\centering
	\includegraphics[width=0.9\columnwidth]{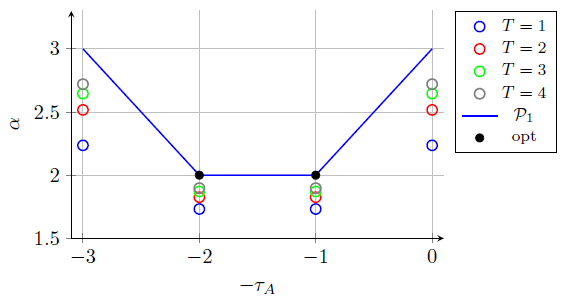}
	\caption{Gains as as function of the acausal delay for $\mathcal{P}_1$ and $\hat\tau_N=3$.}
	\label{fig:alphaT_tauMin_P1_tau_3}
\end{figure}

Evaluating \eqref{eq:alpha} with \eqref{eq:wk_P1} yields $\alpha_T$ as depicted in Fig.~\ref{fig:alphaT_T_P1_tauMax_3} for different acausal delays $\tau_A$ (blue plus signs). 
The black circles are the resulting worst case gains found numerically by direct variation of all possible combinations of packet delays $\tau_j$. Hence, one can mathematically reproduce the true worst case gains $\alpha_T$ that tend to $\alpha=\bar\tau$ for $T\rightarrow\infty$, see dashed lines in Fig.~\ref{fig:alphaT_T_P1_tauMax_3}.
An alternative way to plot the results from Fig.~\ref{fig:alphaT_T_P1_tauMax_3} is presented in Fig.~\ref{fig:alphaT_tauMin_P1_tau_3}, where circles indicate the different values for $\alpha_T$ for different $T$ and the blue solid line visualizes $\alpha$ as a function of $\tau_A$.
As a consequence, the optimal choice for the acausal delay (black dots) is \eqref{eq:th_tauA_P1}
and $\alpha^\ast$ equal to \eqref{eq:th_tauA_alpha_P1}, see Theorem~\ref{th:NCS_stability}. The resulting worst case delay pattern is
\begin{equation}
\tau_j = \begin{cases}
\ \ \,\bar\tau^\ast	   & \text{if}\quad \bar\tau^\ast = \hat\tau_N-\tau_A \\
-\bar\tau^\ast   & \text{otherwise}
\end{cases} \, .
\end{equation}

\subsection{Protocol $\mathcal{P}_3$}

The calculation of $\alpha^\ast$ is more demanding for protocol $\mathcal{P}_3$, because the delay patterns from the previous section only partially reproduce gains $\alpha_T$, e.g.,  for $\mathcal{P}_3$ and $\tau_A=\hat\tau_N$. This is visualized in Fig.~\ref{fig:alphaT_T_P3_tauMax_3_P1P3aP3bP3c} using blue plus signs. An analysis of the delay patterns that correspond to the numerically found worst case gains $\alpha_T$ reveals that three different delay patterns $\mathcal{P}_3^{\prime}$, $\mathcal{P}_3^{\prime\prime}$ and $\mathcal{P}_3^{\prime\prime\prime}$ have to be considered to mathematically describe the gains related to protocol $\mathcal{P}_3$.

\subsubsection{Delay Pattern $\mathcal{P}_3^{\prime}$}

\begin{figure}[!t]\centering
	\includegraphics[width=\columnwidth]{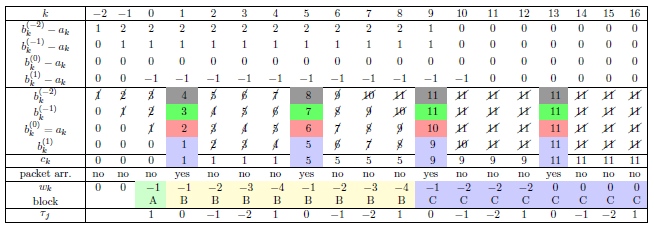}
	\caption{Packet pattern and inner signals of the uncertainty for $\mathcal{P}_3^{\prime}$, $\hat\tau_N=3$, $T=10$ and $\tau_A=2$.}
	\label{fig:P3_1_tauMax3_-2_1}
\end{figure}

\begin{figure}[!t]\centering
	\includegraphics[width=\columnwidth]{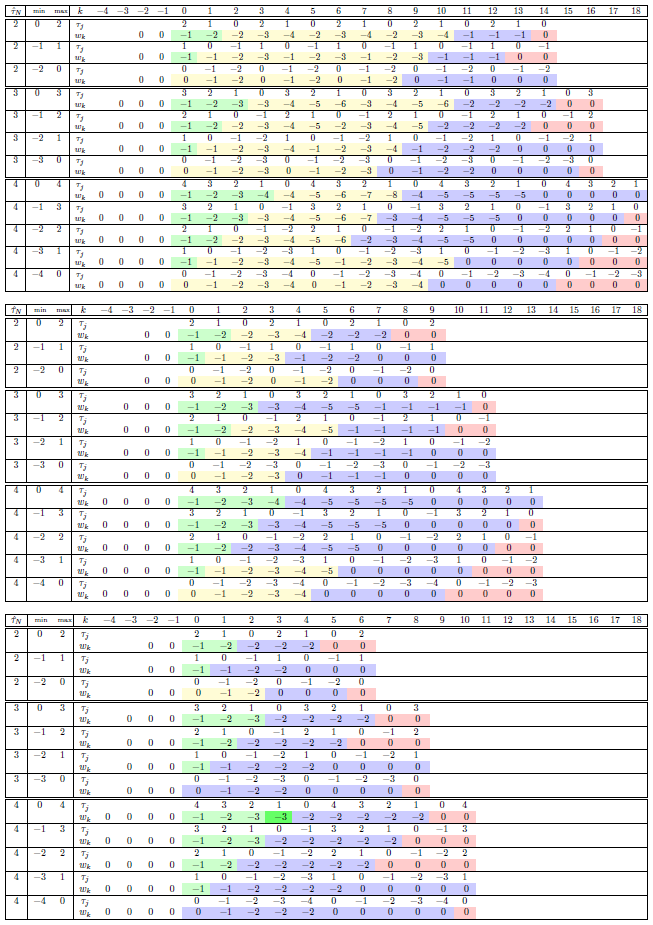}
	\caption{Resulting delay patterns $(\tau_j)$ and worst case output sequences $(w_k)$ for $\mathcal{P}_3^{\prime}$, $\hat\tau_N\in\{2,3,4\}$, $T=10$ (top), $T=5$ (middle), and $T=2$ (bottom).}
	\label{fig:P3_1_T10_T2}
\end{figure}

The first packet arrival pattern is based on \cite{Steinberger2020_arxiv} for the causal case of $\mathcal{P}_3$. Figure~\ref{fig:P3_1_tauMax3_-2_1} show the associated delay pattern, inner signals of the uncertainty and the worst case output sequence for $\hat\tau_N=3$, $T=10$ and $\tau_A=2$.
This is then generalized for different $\hat\tau_N$, $T$, and $\tau_A$ as shown in Fig.~\ref{fig:P3_1_T10_T2} for $\hat\tau_N\in\{2,3,4\}$ and $T\in\{2,5,10\}$.
Hence, delay pattern
\begin{equation}
	(\tau_j) = \big( \underbrace{\left(\hat\tau_N-\tau_A\right), \left(\hat\tau_N-\tau_A-1\right), \ldots, -\tau_A}_{\star}, \star, \star, \cdots\big)
\end{equation}
yields $||(w_k)_T||_2^2 = A + B + C + D$ consisting of four parts A, B, C and D. Introductory part A consists of $\hat\tau_N-\tau_A$ samples. It contributes to $||(w_k)_T||_2^2$ with
\begin{equation}
	A = \sum_{i=1}^{\hat\tau_N-\tau_A} \Big(\min\big\{ i, T+1\big\} \Big)^2 \bar v^2 
\end{equation}
and $T\ge0$, where this formulation with the min-operator is used to account for cases where $T+1<\hat\tau_N$ as, e.g., shown in the Fig.~\ref{fig:P3_1_T10_T2} for $T=2$, $\hat\tau_N=4$ and $\tau_A=0$. Block B consists of $k_1$ (yellow) blocks of length $\hat\tau_N+1$, containing a repeating sequence $\big(\hat\tau_N-\tau_A, \hat\tau_N-\tau_A+1, \ldots, ... , 2\hat\tau_N-\tau_A\big)$ with a squared norm of
\begin{equation}\label{eq:d}
	d = \sum_{i=0}^{\hat\tau_N} \big(\hat\tau_N-\tau_A + i\big)^2 \bar v^2
\end{equation}

The causal case in \cite{Steinberger2020_arxiv} provides the basis to generalize the calculation of $k_1$ for the acausal one. Overall $T+2\hat\tau_N+1$ are are taken into account in the analysis, where block B contribute with $\hat\tau_N-\tau_A$ samples. To get $k_1$, one subtracts the length of the repeating sequence $(\hat\tau_N+1)$ from the remaining $T+1+\hat\tau_N+\tau_A$ samples so many times that $\le 3\hat\tau_N$ samples are left. The remaining samples $k_2 = T+1+\hat\tau_N+\tau_A - k_1 (\hat\tau_N+1)$ are split into 
\begin{equation}
	k_3 = \bigg\lfloor \frac{k_2}{\hat\tau_N+1} \bigg\rfloor
\end{equation}
times $\hat\tau_N+1$ samples (block C, blue) and the left $k_2 - k_3 (\hat\tau_N+1)$ samples shown in red. 
This yield
\begin{align}\label{eq:wk_P3a}
	&\big|\big|(w_k)_T\big|\big|_2 = \underbrace{\sum_{i=1}^{\hat\tau_N-\tau_A} i^2 \bar v^2}_A + \underbrace{k_1  \sum_{i=0}^{\hat\tau_N} \big(\hat\tau_N-\tau_A + i\big)^2 \bar v^2}_B + 
\end{align}
\begin{align}
	&\quad \underbrace{\sum_{j=k_1}^{k_1+k_3-1} \Bigg\{ \sum_{i=\hat\tau_N-\tau_A+j(\hat\tau_N+1)}^{2\hat\tau_N-\tau_A+j(\hat\tau_N+1)} \left(a_i - a_{j(\hat\tau_N+1)}\right)^2\Bigg\}}_C + \nonumber\\
	&\quad \underbrace{\sum_{i=\hat\tau_N-\tau_A+(k_1+k_3)(\hat\tau_N+1)}^{T+2\hat\tau_N} \left( a_i - a_{\hat\tau_N-\tau_A+(k_1+k_3)(\hat\tau_N+1)} \right)^2}_D \nonumber , 
\end{align}
which is used in the calculation of $\alpha_T$. The $\ell_2$ gain follows
from \eqref{eq:alpha}, \eqref{eq:d}, \eqref{eq:wk_P3a} and 
\begin{equation}
	\alpha_T^2 = \frac{A + \frac{T+1}{\hat\tau_N+1} d +C+D}{1+T} = \frac{A+C+D}{1+T} + \frac{d}{\hat\tau_N+1}
\end{equation}
for $T\rightarrow\infty$ such that 
\begin{align}
	d &= \sum_{i=1}^{\hat\tau_N} \big(\Delta\tau - 1 + i\big)^2 \bar v^2 \\
	&= \bigg[(\Delta\tau-1)^2 \sum_{i=1}^{\hat\tau_N+1} 1 + 2 (\Delta\tau-1) \sum_{i=1}^{\hat\tau_N+1} i + \sum_{i=1}^{\hat\tau_N+1} i^2  \bigg] \bar v^2 \nonumber\\
	&= (\hat\tau_N+1) \bigg[\Delta\tau^2 + \Delta\tau\,\hat\tau_N + \frac{1}{3} \hat\tau_N^2 + \frac{1}{6} \hat\tau_N\bigg] \bar v^2 \nonumber
\end{align}
with $\Delta\tau = \hat\tau_N-\tau_A$ and so 
\begin{equation}\label{eq:alpha_P1a}
	\alpha_{\mathcal{P}_3^{\prime}} = \sqrt{\Delta\tau^2 + \Delta\tau\,\hat\tau_N + \frac{1}{3} \hat\tau_N^2 + \frac{1}{6} \hat\tau_N} \ .
\end{equation}

Figure~\ref{fig:alphaT_T_P3_tauMax_3_P1P3aP3bP3c} shows the results for $\alpha_T$ and the corresponding $\ell_2$ gains $\alpha_{\mathcal{P}_3^{\prime}}$ for different acausal delays. Condition \eqref{eq:alpha_P1a} reduces to $\alpha=\sqrt{\frac{\bar\tau}{6}(14 \bar\tau+1)}$ in causal case where $\Delta\tau=\hat\tau_N=\bar\tau$, see \cite{Steinberger2020_arxiv}.
The presented delay pattern allows to reproduce additional points in Fig.~\ref{fig:alphaT_T_P3_tauMax_3_P1P3aP3bP3c} as, e.g., for $\tau_A=0$ and $\tau_A=1$.

\subsubsection{Delay Pattern $\mathcal{P}_3^{\prime\prime}$}

\begin{figure}[!t]\centering
	\includegraphics[width=\columnwidth]{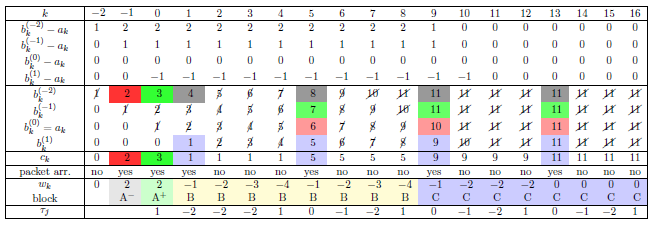}
	\caption{Packet pattern and inner signals of the uncertainty for $\mathcal{P}_3^{\prime\prime}$, $\hat\tau_N=3$, $T=10$ and $\tau_A=2$.}
	\label{fig:P3_2_tauMax3_-2_1}
\end{figure}

\begin{figure}[!t]\centering
	\includegraphics[width=\columnwidth]{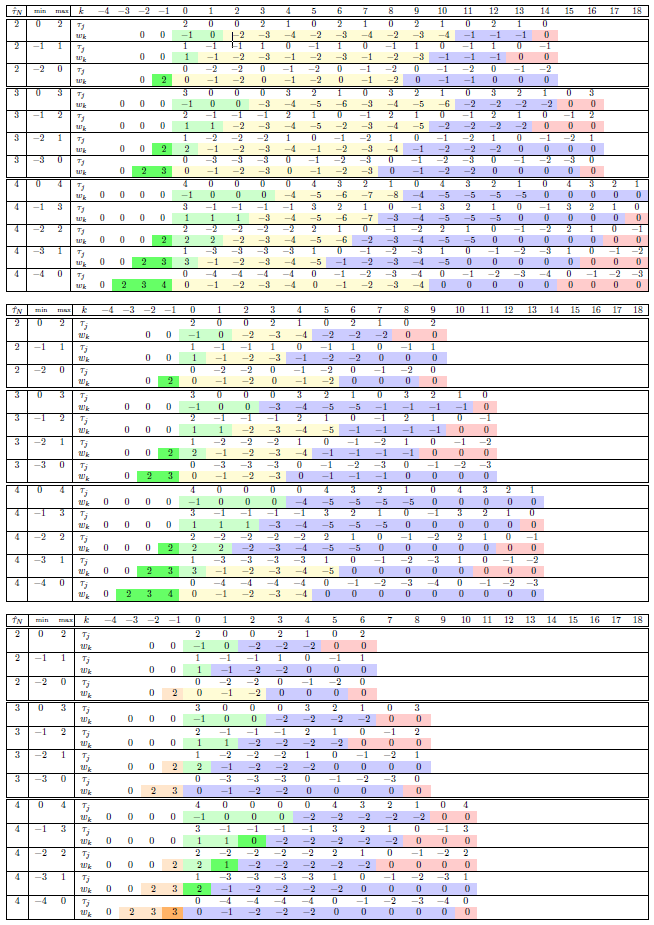}
	\caption{Resulting delay patterns $(\tau_j)$ and worst case output sequences $(w_k)$ for $\mathcal{P}_3^{\prime\prime}$, $\hat\tau_N\in\{2,3,4\}$, $T=10$ (top), $T=5$ (middle), and $T=2$ (bottom).}
	\label{fig:P3_2_T10_T2}
\end{figure}

A second delay pattern follows by modifying the introductory part of $\mathcal{P}_3^{\prime}$ such that
\begin{align}
	(\tau_j) &= \big((\hat\tau_N-\tau_A), \underbrace{-\tau_A, -\tau_A, \ldots, -\tau_A}_{\hat\tau_N \text{ times}}, \\
	&\quad\ \underbrace{\left(\hat\tau_N-\tau_A\right), \left(\hat\tau_N-\tau_A-1\right), \ldots, -\tau_A}_{\star}, \star, \star, \cdots\big) \nonumber .
\end{align}
The resulting packet patterns and inner signals of the uncertainty are exemplified in Fig.~\ref{fig:P3_2_tauMax3_-2_1} for $\hat\tau_N=3$, $T=10$ and an acausal delay of $\tau_A=2$.
Figure~\ref{fig:P3_2_T10_T2} illustrate the patterns for different $\hat\tau_N$, $\tau_A$ and truncations $T$.
In contrast to \eqref{eq:wk_P3a}, the introductory part A is split into $A^+$ for $k\ge0$ and $A^-$ for $k<0$. All other parts remain the same as in $\mathcal{P}_3^{\prime}$. According to the tables in Fig.~\ref{fig:P3_2_T10_T2}, the modified intro can mathematically be described by
\begin{subequations}\label{eq:A_P3b}
\begin{align}
	A^+ &= \begin{cases}
		\bar v^2 \qquad\quad \text{if} \quad \tau_A=0\\
	\displaystyle\sum_{i=1}^{\hat\tau_N-\tau_A} \bigg(\min\Big\{T, \tau_A, \max\big\{0, T-i+1\big\} \Big\} \bigg)^2 \bar v^2 \ , \\
	\qquad\qquad \text{otherwise}
	\end{cases}
\end{align}
\begin{align}
	A^- &= \begin{cases}
		0 & \text{if}\quad \tau_A\in\{0,1\}\\
	\displaystyle \sum_{i=2}^{\tau_A} \Big(\min\big\{i, T+1\big\}\Big)^2 \bar v^2 & \text{otherwise}
	\end{cases} \, .
\end{align}
\end{subequations}
Relation~\eqref{eq:alpha} in combination with \eqref{eq:wk_P3a} and \eqref{eq:A_P3b}
allows to calculate $\alpha_T$ as presented in Fig.~\ref{fig:alphaT_T_P3_tauMax_3_P1P3aP3bP3c}. It can be observed that the limit of $\alpha_T$ for $T\rightarrow\infty$ is equal to $\mathcal{P}_3^{\prime}$. However, larger vales 
\begin{equation}
	\alpha_{\mathcal{P}_3^{\prime\prime}} = \sup_T \alpha_{T,\mathcal{P}_3^{\prime\prime}}	
\end{equation}
result for smaller $T$, as, e.g. for $\tau_A=2$ and $T=3$ in Fig.~\ref{fig:alphaT_T_P3_tauMax_3_P1P3aP3bP3c}.

\subsubsection{Delay Pattern $\mathcal{P}_3^{\prime\prime\prime}$}

\begin{figure}[!t]\centering
	\includegraphics[width=\columnwidth]{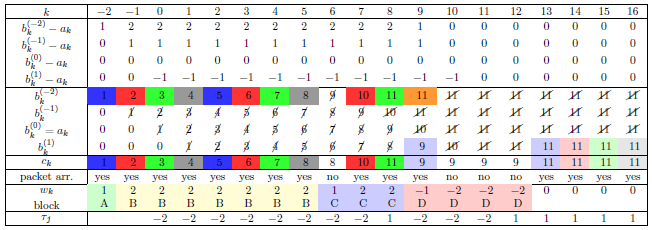}
	\caption{Packet pattern and inner signals of the uncertainty for $\mathcal{P}_3^{\prime\prime\prime}$, $\hat\tau_N=3$, $T=10$ and $\tau_A=2$.}
	\label{fig:P3_3_tauMax3_-2_1}
\end{figure}

\begin{figure}[!t]\centering
	\includegraphics[width=\columnwidth]{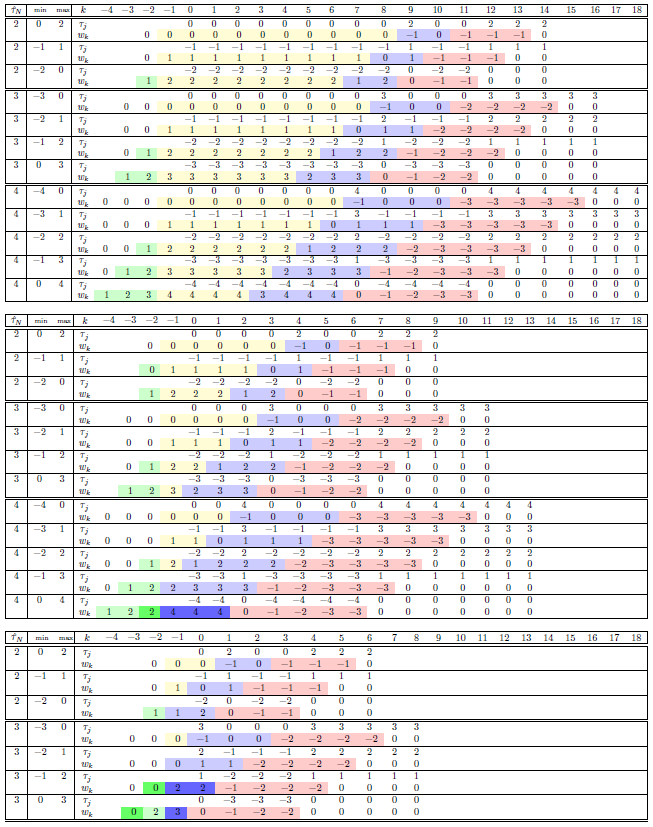}
	\caption{Resulting delay patterns $(\tau_j)$ and worst case output sequences $(w_k)$ for $\mathcal{P}_3^{\prime\prime\prime}$, $\hat\tau_N\in\{2,3,4\}$, $T=10$ (top), $T=5$ (middle), and $T=2$ (bottom).}
	\label{fig:P3_3_T10_T2}
\end{figure}

\begin{figure}[!t]\centering
	\includegraphics[width=\columnwidth]{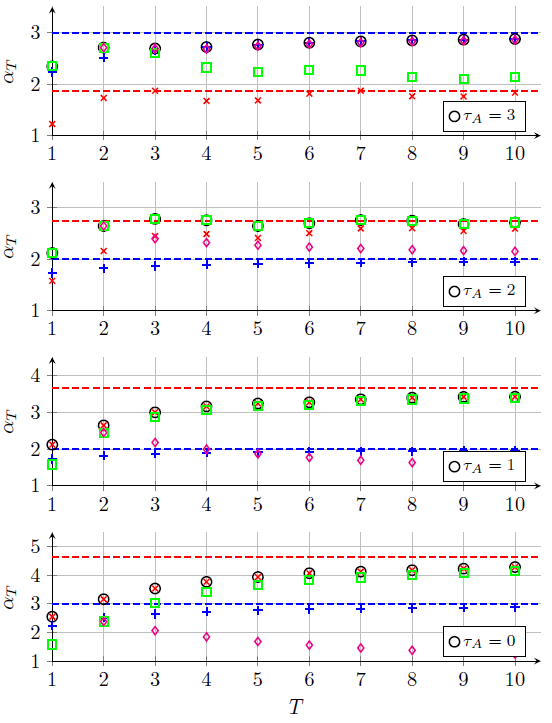}
	\caption{Gains as a function of truncation point $T$ and the acausal delay $\tau_A$ for  $\hat\tau_N=3$ and $\mathcal{P}_1$ (blue), $\mathcal{P}_3^{\prime}$ (red), $\mathcal{P}_3^{\prime\prime}$ (green), $\mathcal{P}_3^{\prime\prime\prime}$ (magenta).}
	\label{fig:alphaT_T_P3_tauMax_3_P1P3aP3bP3c}
\end{figure}

Almost all points in Fig.~\ref{fig:alphaT_T_P3_tauMax_3_P1P3aP3bP3c} can be mathematically described by using either  $\mathcal{P}_1$, $\mathcal{P}_3^{\prime}$ or $\mathcal{P}_3^{\prime\prime}$. However, this is not true for, e.g., $\tau_A=3$ and $T=3$.
Thus, the additional delay pattern
\begin{align}\label{eq:tau_P3c}
	(\tau_j) &= \big(\underbrace{-\tau_A, -\tau_A, \ldots, -\tau_A}_{T-\hat\tau_N+1 \text{ elements}}, (\hat\tau_N-\tau_A) , \\
	&\quad\ \ \underbrace{-\tau_A, -\tau_A, \ldots, -\tau_A}_{\hat\tau_N \text{ elements}}, (\hat\tau_N-\tau_A), (\hat\tau_N-\tau_A), \ldots\big) \nonumber
\end{align}
for $T-\hat\tau_N+1 \ge 0$, i.e. $T \ge \hat\tau_N-1$ is utilized. The corresponding patterns and inner signals are shown in Fig.~\ref{fig:P3_3_tauMax3_-2_1} and \ref{fig:P3_3_T10_T2}
and for different values of $\hat\tau_N$, $\tau_A$, and $T$. Block A, C and D consist of $\hat\tau_N-1$, $\hat\tau_N$ and $\hat\tau_N+1$ samples, respectively. As a result, the length of (yellow) block B is $T-\hat\tau_N-\tau_A+2$. 
The worst case norm of output sequence $(w_k)$ for $\mathcal{P}_3^{\prime\prime\prime}$ is given by
\begin{equation}
	||w_k||_{2,T}^2 = A + B + C + D
\end{equation}
with
\begin{subequations}\label{eq:ABCD_P3c}
\begin{align}
	A &= \begin{cases}
		0 & \text{if}\quad \tau_A\in\{0,1\}\\
		\displaystyle\sum_{i=1}^{\tau_A-1} \Big(i-\delta^{(i)} \Big)^2 \bar v^2 & \text{otherwise}
	\end{cases}
\end{align}
\begin{align}
	\delta^{(i)} &= \begin{cases}
		1 & \text{if}\quad i=(T-\hat\tau_N+2) \ \wedge\  (T<\hat\tau_N-2+\tau_A)\\
		0 & \text{otherwise}
	\end{cases}\\
	B &= \begin{cases}
		\big(T-\hat\tau_N-\tau_A+2\big) \tau_A^2 \bar v^2 & \text{if}\quad T \ge \hat\tau_N-2+\tau_A\\
		\big(T-\tau_A+2\big) \tau_A^2 \bar v^2 & \text{otherwise}
	\end{cases}\\
	C &= \begin{cases}
		\displaystyle\sum_{i=\tau_A-1}^{\tau_A+\hat\tau_N-2} \Big(\min\big\{i,\tau_A\big\}\Big)^2 \bar v^2 & \text{if}\quad T \ge \hat\tau_N-2+\tau_A\\
		0 & \text{otherwise}
	\end{cases}\\
	D &= \sum_{i=\hat\tau_N-\tau_A}^{2\hat\tau_N-\tau_A} \Big(\min\big\{i,\hat\tau_N-1\big\}\Big)^2 \bar  v^2
\end{align}
\end{subequations}
for $T \ge \max\{\hat\tau_N-1\}$. The different cases in \eqref{eq:ABCD_P3c} allow to
correctly calculate $\alpha_{T,\mathcal{P}_3^{\prime\prime\prime}}$ 
also for small truncation $T$ as shown in Fig.~\ref{fig:P3_3_T10_T2} as well as
\begin{equation}
	\alpha_{\mathcal{P}_3^{\prime\prime\prime}} = \sup_T \alpha_{T,\mathcal{P}_3^{\prime\prime\prime}}	\ .
\end{equation}

\subsubsection{Protocol $\mathcal{P}_3$}

\begin{figure}[!h]\centering
	\includegraphics[width=0.9\columnwidth]{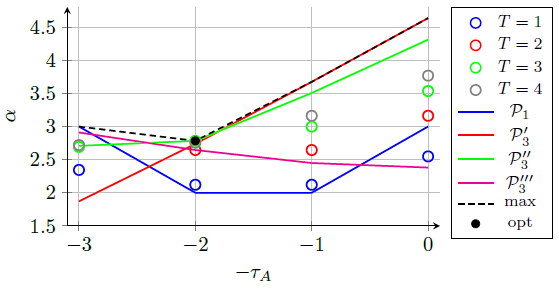}
	\caption{Gains as as function of the acausal delay for $\mathcal{P}_3$ and $\hat\tau_N=3$}
	\label{fig:alpha_tauMin_P3_tau_3}
\end{figure}

\begin{figure}[!h]\centering
	\includegraphics[width=0.9\columnwidth]{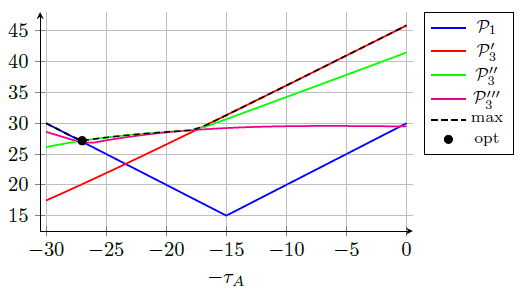}
	\caption{Gains as as function of the acausal delay for $\mathcal{P}_3$ and  $\hat\tau_N=30$}
	\label{fig:alpha_tauMin_P3_tau_30}
\end{figure}

To find the overall finite $\ell_2$ gain for protocol $\mathcal{P}_3$ one has to combine $\mathcal{P}_1$, $\mathcal{P}_3^{\prime}$, $\mathcal{P}_3^{\prime\prime}$ and $\mathcal{P}_3^{\prime\prime\prime}$. Figures~\ref{fig:alpha_tauMin_P3_tau_3} and \ref{fig:alpha_tauMin_P3_tau_30} visualize the results for the gain calculations for  maximal admissible time-varying delays of $\hat\tau_N=3$ and $\hat\tau_N=30$.

Different gains that depend on the actual choice of the acausal delay $\tau_A$ follow for the considered delay patterns. Values for $\alpha_T$ that are found by direct variation all possible combinations of packet delays $\tau_j$ are indicated as black circles in Fig.~\ref{fig:alphaT_T_P3_tauMax_3_P1P3aP3bP3c} to show that the used mathematical description exactly reproduces all $\alpha_T$. In Fig.~\ref{fig:alpha_tauMin_P3_tau_3}, the results for $T=\{1,2,3,4\}$ are shown depending on acausal delay $\tau_A$. Results for larger $T$ are omitted for clarity in the plot. The optimal $\ell_2$ gain $\alpha^\ast$ and the optimal acausal delay $\tau_A^\ast$ are found numerically as the minimum among the different patterns (black dots in Fig.~\ref{fig:alpha_tauMin_P3_tau_3} and \ref{fig:alpha_tauMin_P3_tau_30}) such that
\begin{equation}\label{eq:alpha_P3}
	\alpha^\ast = \min_{\tau_A}\bigg\{\max_{\tau_A}\Big\{ \alpha_{\mathcal{P}_1},\alpha_{\mathcal{P}_3^{\prime}},\alpha_{\mathcal{P}_3^{\prime\prime}},\alpha_{\mathcal{P}_3^{\prime\prime\prime}} \Big\}\bigg\} \, ,
\end{equation}
with the corresponding optimal choice for the acausal delay
\begin{equation}\label{eq:tau_P3}
	\tau_A^\ast = \argmin_{\tau_A}\bigg\{\max_{\tau_A}\Big\{ \alpha_{\mathcal{P}_1},\alpha_{\mathcal{P}_3^{\prime}},\alpha_{\mathcal{P}_3^{\prime\prime}},\alpha_{\mathcal{P}_3^{\prime\prime\prime}} \Big\}\bigg\} \, .
\end{equation}
The numeric evaluation of \eqref{eq:alpha_P3}, \eqref{eq:tau_P3} results in relation \eqref{eq:th_alpha_P3} in Theorem~\ref{th:NCS_stability} and Table~\ref{tab:P3_optimal_values}.

As pointed out in Remark~\ref{re:overest}, it is possible to make use of the gain that follows for $\mathcal{P}_1$ evaluated at $\tau_A=\hat\tau_N$ as an over-estimation of $\alpha^\ast$ for protocol $\mathcal{P}_3$, cf. Fig~\ref{fig:alpha_tauMin_P3_tau_3} and \ref{fig:alpha_tauMin_P3_tau_30}.


\bibliographystyle{IEEEtran}
\bibliography{SGT_NCS_acausal_arxiv}

\end{document}